\documentclass[10pt,twocolumn]{article}
\usepackage[draft]{hyperref}
\usepackage{amsmath}
\usepackage{graphicx}
\usepackage{stfloats}
\usepackage{placeins}
\usepackage{fullpage}
\usepackage{fixltx2e}
\usepackage[superscript,biblabel]{cite}

    \makeatletter
    \let\@fnsymbol\@arabic %
    \makeatother
    \newcommand{\ignore}[1]{}
\begin{document}

\title{Quantum correlation enhanced super-resolution localization microscopy enabled by a fiber bundle camera}

\author{Yonatan Israel\footnotemark[1]\;,
            Ron Tenne\footnotemark[1]\;,
            Dan Oron\footnotemark[1]\;\;\&
            Yaron Silberberg
            \thanks{Department of Physics of Complex Systems, Weizmann Institute of Science, Rehovot 76100, Israel.}}

\twocolumn[
\maketitle
  \begin{@twocolumnfalse}
\abstract{ 
\noindent Despite advances in low-light level detection\ignore{\cite{SandoghdarReview15}}, single-photon methods such as photon correlation\ignore{\cite{orrit2002review,SandogdharPRL2010}} have rarely been used in the context of imaging. The few demonstrations, for example of sub-diffraction limited imaging utilizing quantum statistics of photons\ignore{\cite{CuiPRL2013,OsipNanoletters2013}}, have remained in the realm of proof-of-principle demonstrations. This is primarily due to a combination of low values of fill factors, quantum efficiencies, frame rates and signal-to-noise characteristic of most available single-photon sensitive imaging detectors\ignore{\cite{SPcams2014}}. Here we describe an imaging device based on a fiber bundle coupled to single-photon avalanche detectors, which combines a large fill factor, a high quantum efficiency, a low noise and scalable architecture. Our device enables localization based super-resolution microscopy in a non-sparse non-stationary scene, utilizing information on the number of active emitters, as gathered from non-classical photon statistics.}
\vspace{3mm}
  \end{@twocolumnfalse}
]

\vspace{3mm}
\noindent{\large\textbf{Introduction}}

\noindent Far-field optical microscopy, an important workhorse in biological research, is fundamentally limited by diffraction, as was established by Abbe\cite{abbe1873} and Rayleigh\cite{Rayleigh1896}. The attainable resolution is therefore limited to approximately half the wavelength of light. In the past two decades, several successful schemes to overcome the diffraction limit in microscopy were developed\cite{STED_PNAS_2000,rust2006sub,betzig2006imaging,dertinger2009fast,SandoghdarReview15}. Many of these utilize the concept of precise localization of a single emitter in a time series of sparse frames\cite{localization_review_NMeth2014}.
One inherent problem of these methods is the sparsity requirement, that is a single emitter per diffraction limited spot per frame at most, slowing down the acquisition of super-resolved images\cite{Small_biophys2008}.
Several schemes for localizing multiple emitters were already presented\cite{Huang2011_multi,Zhu2012NatMehod,sCMOS_Loc_2013,TonyICA_APL13}; however, these algorithms yield limited performance and lack robustness \cite{wang2012palmer,localization_algorithms_review_NMeth2014}.
As shown in this work, gathering additional information on the number of active emitters, namely, photon correlation statistics, enables localization in non-sparse scenes.

In the last few years, the use of non-classical photon statistics for sub-diffraction limited imaging has been theoretically studied\cite{OsipSR_PRA,Tsang_PhysRevLett2016,CosmoPirandola_PRL2016} and demonstrated by photon correlation measurement in both a widefield\cite{OsipNanoletters2013} and a confocal\cite{CuiPRL2013} imaging geometries. In practice, however, both realizations do not exhibit a viable pathway for super-resolution imaging or particle tracking. Here we propose and demonstrate a method that rather than using photon correlation information directly, utilizes it for multi-emitter localization in a time-dependent scene. By analyzing both the simultaneous detections of photons and spatial information, one can accurately determine the number of emitters contributing to an image\cite{COPS_CPC_2014,MappingSTED_NComm15} and localize them. In contrast with optimization based schemes\cite{Zhu2012NatMehod,TonyICA_APL13}, here experimental information which was previously unavailable is provided as input for the localization algorithm.

\begin{figure} 
\includegraphics[width=\columnwidth]{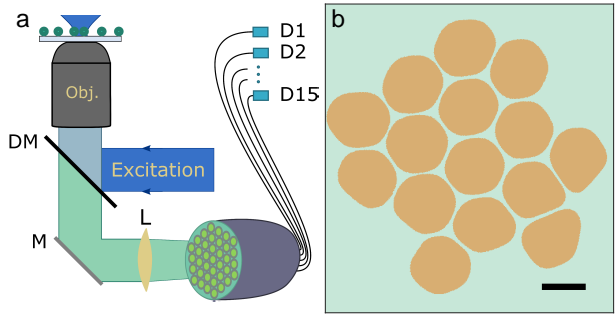}
\caption{\label{fig:setup} {\footnotesize Measuring quantum correlations in a confocal microscope. (a) Schematic of a single-photon fiber bundle camera (SFICAM) with 15 single-photon avalanche detectors (SPADs). (b) Cross section of the fiber bundle. This segmented image was compiled by thresholding an array of optical microscope images in which light was input into one of the fibers (see more details in Supplementary Figure 5). Scale bar, 100$\mu$m.}}
\end{figure}

\vspace{3mm}
\noindent{\large\textbf{Results}}

\noindent\textbf{Principle}

\noindent A key ingredient required for realization of this scheme is a fast, low-noise, single-photon sensitive imaging detector. Over the last two decades, progress in the technology of low-light level sensitive cameras, was an important enabling factor in the development of super-resolution microscopy techniques\cite{AccuracityEMCCDsNMeth13,sCMOS_Loc_2013}. Still, they are quite noisy, and the frame rate of such cameras is limited to about 1 kHz\cite{SPcams2014}, washing out information contained at higher temporal frequencies. Alternative detectors based on integrated single-photon avalanche detectors (SPAD) arrays on a chip typically suffer from very low fill factors, even when using microlenses\cite{microlens2014}.

Our imaging device, the single-photon fiber bundle camera (SFICAM), is a low pixel-number camera, constructed from a fiber bundle, in which each fiber acts as a pixel and guides photons to a SPAD, as shown in Fig. \ref{fig:setup} (for more details see Methods). This device combines spatial information with a single-photon sensitivity and nanosecond scale temporal resolution, capable of detecting emission transients orders of magnitude faster than the $1\,$ms temporal resolution of typical cameras. Since the detectors are separated from the imaging facet of the bundle, a fill factor of over 80\% is achieved (see Supplementary Note 1 and Supplementary Figure 1). These characteristics allow us to efficiently analyze quantum photon-photon correlations within an image.

Many fluorophores are inherently single-photon emitters, for example dye molecules and quantum dots (QDs)\cite{OrritPh  otonSource2005}. Therefore, simultaneously detected pairs of photons from such fluorophores provide valuable information concerning the number of emitters in every frame.
Supposing that $n$ identical single-photon emitters are measured, their zero delay ($\tau=0$) second-order photon correlation\cite{IntroQuantOPt} ($g^{(2)}$) will be
\begin{align}\label{eq:g2}
\centering
    g^{(2)}(\tau=0) = 1 - \frac{1}{n}.
\end{align}
By measuring quantum correlations, the number of active emitters can be found, as seen from equation (\ref{eq:g2}). In particular, it can determined from such measurements whether only a single emitter is switched on in the detection volume. We therefore continuously evaluate $g^{(2)}(0)$ in order to estimate the number of emitters contributing to an image at every point along the acquisition time.
Finally, a localization algorithm can be applied to localize the emitters, using the precise number of emitters in the image.

\begin{figure}
  \includegraphics[width=\columnwidth]{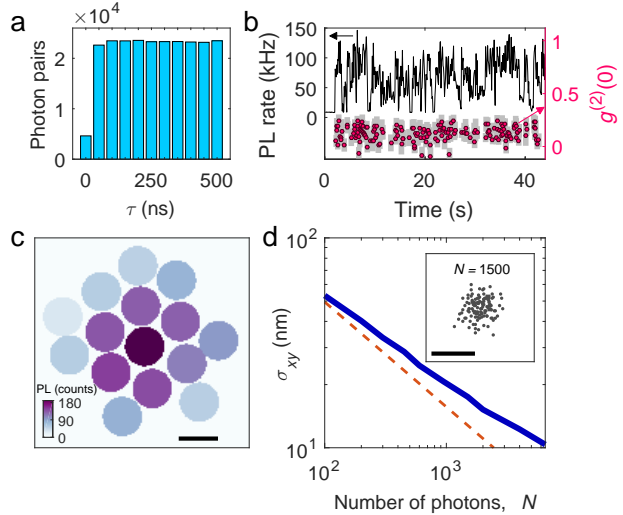}
\caption{\label{fig:1QD} {\footnotesize Localization of a single emitter using a fiber bundle. (a) Two-photon correlation count, and (b) Photoluminescence (PL) count rate (black line), summed over all detectors, and quantum correlation (red dots), summed over all detector pairs, for a single quantum dot (QD). Error bars represent $\pm\sigma$ statistical error (gray). (c) A single QD imaged by the bundle camera using $N=1500$ photons in $15$ ms. Color bar represent PL count. Scale bar, 100 $\mu$m. (d) Two-dimensional localization error $\sigma_{xy}$ measured for a single QD (solid blue) and theoretical precision (see text). Inset shows 200 localizations using $N=1500$ photons for a single QD, where the localization precision is $\sigma_{xy}=$ 20 nm. Scale bar, 100$\mu$m.
}}
\end{figure}
\vspace{3mm}

\noindent\textbf{Single emitters}

\noindent Fig. \ref{fig:1QD} presents a typical measurement of a single QD emitter. A photon-correlation measurement, commonly performed in single-particle spectroscopy experiments, is shown in Fig. \ref{fig:1QD}a. Photon antibunching, manifested by a full dip in the autocorrelation function of the photon stream at zero delay ($\tau = 0$), ensures that indeed the fluorescence source is a single QD emitter. Photoluminescence (PL) rates shown along with $g^{(2)}(0)$ time traces (Fig. \ref{fig:1QD}b) illustrate that the value of $g^{(2)}(0)$ remains constant about zero throughout the entire measurement time whereas the PL fluctuates due to blinking. This comparison highlights the advantage of using the stable photon-correlation signal versus using the fluctuating PL intensity signal for estimating the number of emitters in a scene.
Single-particle tracking can be performed on the same photon trace in order to localize the emitter and analyze localization precision. We apply a least squares minimization algorithm to fit the position of an emitter for an $N = 1500$ photons image (Fig. \ref{fig:1QD}c) using a Gaussian point-spread-function (PSF).
A standard error for two-dimensional localization $\sigma_{xy}^2 = \sigma_x^2+\sigma_y^2$ (where $\sigma_i$ is the standard error for localization along axis $i$) is calculated using consecutive localizations of a single QD over 6 s, to test the precision of the localization procedure\cite{localization_review_NMeth2014}. We compare this precision to a theoretical model\cite{mortensenNMeth2010} accounting for the pixel size and background counts (Fig. \ref{fig:1QD}d) which follows shot-noise scaling. The localization error of our system departs from this model at high $N$, possibly due to mechanical drift of the sample. A drift on the scale of a 100 nm in 50 s of measurement was analyzed for several scenes of particles, and no individual motion of the particles was resolved (Supplementary Figure 2).

\begin{figure}
  \includegraphics[width=\columnwidth]{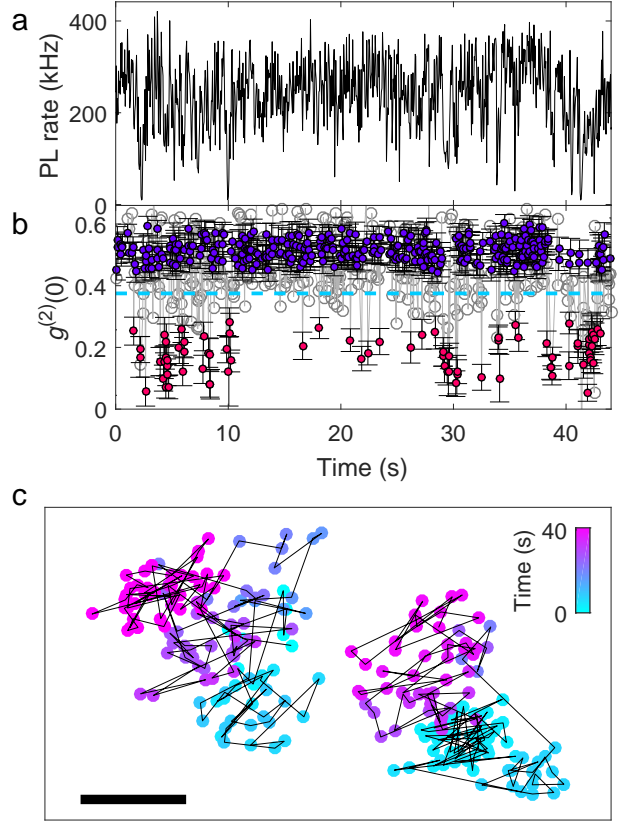}
\caption{\label{fig:2QDs} {\footnotesize Super-resolution localization and single particle tracking using quantum correlations of two emitters undergoing drift. (a) Photon count and (b) quantum correlation for two quantum dots (QDs). Blinking of one QD results in antibunching, $g^{(2)}(0)\approx 0$, shown in red. Instances where more than one emitter is blinked on (blue circles) and where photon statistics is insufficient to count emitters (gray circles) are shown. Error bars represent $\pm\sigma$ statistical error. (c) Single particle tracking (SPT) by repeated localization of two QDs. The separation between the QDs is about 100 nm. Scale bar, 50 nm.
}}
\end{figure}

\vspace{3mm}
\noindent\textbf{Two emitters}

\noindent By using the localization precision together with the extra information provided by photon correlations we demonstrate super-resolved tracking through an analysis algorithm outlined below and detailed in Supplementary Note 2. The Photoluminescence (PL) rates ((Fig. \ref{fig:2QDs}a)) shown along the second-order photon correlation function $g^{(2)}(\tau)$ (Fig. \ref{fig:2QDs}b) are measured and analyzed in time bins of 0.1 s.
To perform single-emitter localization, we post-select time bins in which only a single emitter was blinked on by thresholding the value of $g^{(2)}(0)$ below $0.375$ (red circles). Instances in which more than one emitter is blinked on, having a higher photon correlation $g^{(2)}(0)$ value (blue circles), are rejected by the single-emitter algorithm, as well as those with insufficient photon statistics (clear gray circles). Selecting a threshold of $g^{(2)}(0)=0.375$ rather than the $0.5$ value, inferred from equation (\ref{eq:g2}) for $n=2$ emitters, takes into account some dispersion of PL intensities within the QDs ensemble (a detailed derivation of the single particle criterion can be found in Supplementary Notes 3 and 4).

Localization and tracking of two emitters using our single emitter localization algorithm for segments composed of $N=1500$ photons are shown in Fig. \ref{fig:2QDs}c. The two emitters separated by about 100 nm, are clearly distinguishable as they move toward the top and left corner of the image. Additional examples of distinguishing two emitters with sub-wavelength separation are found in Supplementary Figure 3.
The imaging resolution in this case is given by the single emitter localization precision $\sigma_{xy}$, measured to be 20 nm for $N=1500$ photons (Fig. \ref{fig:1QD}d).
In this example the two QDs are immobilized on the glass substrate while their movement is a result of sample drift, accounting for their correlated motion.

\begin{figure*}
\centering
 \includegraphics[width=\textwidth]{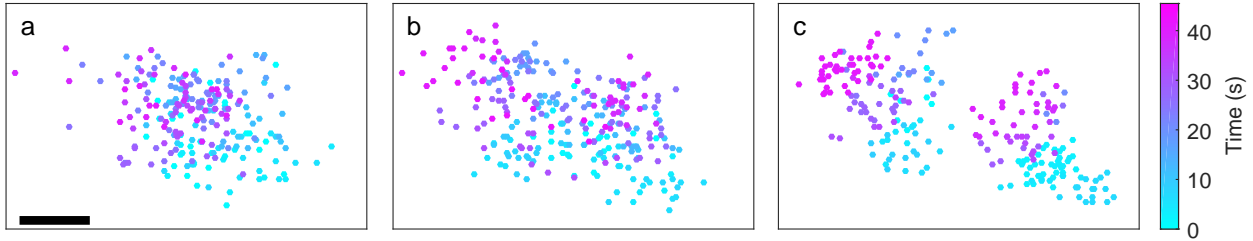}
\caption{\label{fig:2QD_compare} {\footnotesize Localization of two QDs by using (a) all photon counts, (b) a photon count rate threshold of the lowest 10\% intensity periods, and (c) the single emitter anti-bunched photon counts (as in Fig. \ref{fig:2QDs}c). All localizations use $N = 1500$ photons. Scale bar, 50 nm.
}}
\end{figure*}

Figure \ref{fig:2QD_compare} compares post-selection based on quantum-correlations criterion with post-selection according to PL intensity for the localization of a pair of blinking QDs. First, localization without any post-selection, shown in Fig. \ref{fig:2QD_compare}a, results in scattered points that do not resolve the underlying two emitter structure. One might expect that post-selecting localizations from low brightness periods may reveal single emitter events without employing photon correlations. However, the localization scatter from the lowest 10\% intensity periods, shown in Fig. \ref{fig:2QD_compare}b, does not resolve the two emitters. This is due to a significant number of short PL intermittencies resulting in localized points between the two emitters, which obscure the separation of emitter localizations. In contrast, Fig. \ref{fig:2QD_compare}c shows the same data analyzed with the single emitter criterion, clearly resolving the trajectories of the two emitters.

\vspace{3mm}
\noindent\textbf{\large Discussion}

\noindent The SFICAM design allows to image a confocal spot onto an array of a few detectors with a high efficiency.
Scaling up this approach to achieve a high coupling efficiency with multiple detectors in a SPAD arrays\cite{CovaGhioni2011,CharbonFishburn2011} can significantly improve upon their current detection probabilities, limited by low fill-factors (typically $<10\%$). Using a large number of fibers to guide light into a SPAD array would enable single-photon sensitive wide-field imaging with a high temporal resolution. In particular, such a design could be used to extend our technique to perform faster super-resolution localization microscopy in wide-field by making use of quantum correlations.  Furthermore, our optical configuration combines SPADs with conventional confocal microscopy, and could speed-up and enhance the sensitivity of some techniques that image or localize the confocal spot, particularly with confocal super-resolution modalities \cite{STED_PNAS_2000,ISM_PRL10,ConicalSuperRes13,Airyscan2015}.

Our technique relies on two requirements for the emitters: first, they must be single-photon emitting, a requirement met by many fluorophores, including organic dyes and certain fluorescent proteins\cite{DyeAntibunchingNature2000,Review-SRprobes08,singleDyesAndProteinsLocACSNano12,COPS_CPC_2014}.
A second requirement is the detection of a sufficient number of photon correlations events. Common fluorescence microscopy, that uses continuous wave (CW) excitation of organic molecules well below the saturation power of emitters, would typically result in a low number of photon pair detections.  In contrast, a pulsed excitation scheme suppresses photobleaching through the triplet state \cite{donnert2007major,jacques2008enhancing} and therefore allows the use of pulses with almost unity probability of excitation.
In fact, photon correlation measurements with commonly used dye molecule fluorophores were performed under experimental conditions very similar to those used in this work \cite{COPS_CPC_2014}. We emphasize that long term photostability is not pre-requisite from emitters used for super-resolution in our scheme since only 0.1s was enough to extract the valuable $g^{(2)}$ information.



Temporal resolution of super-resolution localization microscopy is limited mainly by the demand of sparse photoswitching\cite{localization_review_NMeth2014,SandoghdarReview15}. Namely, in order to avoid multi-emitter localization events, imaging density should be an order of magnitude lower than one emitter per diffraction limited spot\cite{FALCON2014,SparseLoc2016}. Our approach, integrated into super-resolution localization microscopy, allows to surpass this requirement by precise measurement of the sparsity. By using quantum correlations we measure the number of excited emitters, as follows from equation (\ref{eq:g2}). One can then reject multi-emitter data subsets for single-molecule localizations or even use multi-emitter fitting algorithms\cite{sCMOS_Loc_2013,localization_review_NMeth2014} given the exact number of emitters as extra information.
Our results demonstrate localizations of only two emitters; however, we note that our methods would work for more than two emitters as well. In this case, we note that it would become beneficial to make use of emitters that have faster blinking statistics than the ones used here, to facilitate the occurrence of single-emitter events and localizations of scenes of three emitters or more at viable performance.


Other techniques that achieve super-resolved images using photon and image correlations, show an improvement of the resolution as the square root of the highest order of calculated correlation\cite{dertinger2009fast,OsipNanoletters2013,QC_SPADS_PRL2014}. Practically, signal-to-noise (SNR) and low contrast in high-order correlation pose a limiting factor for such an improvement\cite{dertinger2009fast,OsipNanoletters2013}. In super-resolution optical fluctuation imaging (SOFI) \cite{dertinger2009fast} for example, imaging of two emitters with a five-fold resolution improvement was acquired in several minutes, whereas we demonstrate a ten-fold enhancement in resolution with a temporal resolution of seconds for a small field-of-view.

To summarize, we presented a method that applies quantum photon correlations to accurately localize emitters within a diffraction limited spot. In order to acquire an image together with photon correlations, we utilized a few-pixel confocal camera using a fiber bundle combined with SPADs. Replacing a standard detector of a confocal microscope with the fiber bundle system described above can potentially speed up super-resolved localization microscopy by alleviating the frame sparsity constraint.



\bigskip
\noindent\textbf{\large Methods}

\noindent\footnotesize{\textbf{Microscope setup.} An optical microscope (Zeiss Axiovert 135) is used to image fluorescent samples of QDs. A two-axis piezo stage (P-542.2SL, Physik Instrumente) is used to position the sample. For illumination, a 473nm pulsed picosecond laser diode (Edinburgh Instruments) is used, coupled to a single-mode fiber. The repetition rate of this laser is set to 20MHz.  A $1.4$ numerical aperture objective lens (Plan Apo Vc 100x, Nikon) is used to tightly focus the illuminating laser. The fluorescence is collected by the same objective lens and filtered by dichroic mirrors and filters (FF509-FDi01, SP01-785RS, BLP01-532R, Semrock).
A Galilean beam expander (BE05-10-A, Thorlabs) is placed following the relay lens to magnify the imaged fluorescence spot on to a fiber bundle (A.R.T. Photonics GmbH, Germany). This fiber bundle consists of multimode 100/110 $\mu$m core/clad fibers, fused on one side and fan-out to individual multimode fibers on the other side, and is used to guide photon from an imaged spot to 15 fiber coupled single-photon avalanche photodiodes (SPCM-AQ4C, Perkin-Elemer). For a detailed characterization of the fiber bundle setup see Supplementary Note 4.
The overall detection efficiency of our setup is 12\%, further details about the efficiency are found in the Supplementary Note 1 and Supplementary Table 1.
}

\bigskip
\noindent\footnotesize{\textbf{Data acquisition and analysis.}
A time-correlated single-photon counting board is used for data acquisition in absolute timing mode (DPC-230, Becker \& Hickl GmbH). An excitation pulse trigger is synchronized and recorded at every 40th pulse (0.5MHz).
The correlation analysis and localization algorithms (Supplementary software 1) were implemented in a MATLAB script, post-processing the acquired data. Further details about the algorithms are found in Supplementary Figure 4 and Supplementary Note 5.
}

\bigskip
\noindent\footnotesize{\textbf{QDs and sample preparation.} Samples of CdSe/CdS/ZnS colloidal QDs \cite{OsipNanoletters2013} were prepared by spin coating a low concentration solution mixed with Poly(methyl methacrylate) (PMMA) on a microscope cover-slips. Fluorescence from the these QDs peaks at 610 nm, with a lifetime of 26ns.

\bigskip
\noindent\footnotesize{\textbf{Data availability.}}
The raw data that support the findings of this study are available in figshare repository with the identifier doi:10.6084/m9.figshare.4588723.v1 \cite{rawdata}.

\bigskip
\noindent\textbf{\large Acknowledgements}

\noindent\footnotesize{The authors would like to thank Stella Itzhakov for synthesizing the QDs used in this work and thank Sunil Kumar and Anthony Barsic for fruitful discussions. This work was supported by ERC grant QUAMI, ERC consolidator grant ColloQuanto, the ICore program
of the ISF, the Israeli Nanotechnology FTA program, the Minerva foundation and the Crown Photonics Center.}

\bigskip
\noindent\textbf{\large Author Contributions}

\noindent Y.I. and R.T. equally contributed to this work. All authors contributed to the design of the experiment and preparation of the manuscript. Y.I and R.T. performed the experiment and analyzed the data.

\bigskip
\noindent\textbf{\large Additional Information}

\noindent{Correspondence and requests for materials should be addressed to Y.S. (yaron.silberberg@weizmann.ac.il).}

\bigskip
\noindent\textbf{\large Competing financial interests}

\noindent{The authors declare no competing financial interests.}


\bibliographystyle{naturemag}

\begin{thebibliography}{10}
\expandafter\ifx\csname url\endcsname\relax
  \def\url#1{\texttt{#1}}\fi
\expandafter\ifx\csname urlprefix\endcsname\relax\def\urlprefix{URL }\fi
\providecommand{\bibinfo}[2]{#2}
\providecommand{\eprint}[2][]{\url{#2}}

\bibitem{abbe1873}
\bibinfo{author}{Abbe, E.}
\newblock \bibinfo{title}{Beitr{\"a}ge zur theorie des mikroskops und der
  mikroskopischen wahrnehmung}.
\newblock \emph{\bibinfo{journal}{Archiv f{\"u}r Mikroskopische Anatomie}}
  \textbf{\bibinfo{volume}{9}}, \bibinfo{pages}{413--418}
  (\bibinfo{year}{1873}).

\bibitem{Rayleigh1896}
\bibinfo{author}{Rayleigh, L.}
\newblock \bibinfo{title}{On the theory of optical images, with special
  reference to the microscope}.
\newblock \emph{\bibinfo{journal}{Philosophical Magazine}}
  \textbf{\bibinfo{volume}{42}}, \bibinfo{pages}{167--195}
  (\bibinfo{year}{1896}).

\bibitem{STED_PNAS_2000}
\bibinfo{author}{Klar, T.~A.}, \bibinfo{author}{Jakobs, S.},
  \bibinfo{author}{Dyba, M.}, \bibinfo{author}{Egner, A.} \&
  \bibinfo{author}{Hell, S.~W.}
\newblock \bibinfo{title}{Fluorescence microscopy with diffraction resolution
  barrier broken by stimulated emission}.
\newblock \emph{\bibinfo{journal}{Proceedings of the National Academy of
  Sciences}} \textbf{\bibinfo{volume}{97}}, \bibinfo{pages}{8206--8210}
  (\bibinfo{year}{2000}).

\bibitem{rust2006sub}
\bibinfo{author}{Rust, M.~J.}, \bibinfo{author}{Bates, M.} \&
  \bibinfo{author}{Zhuang, X.}
\newblock \bibinfo{title}{Sub-diffraction-limit imaging by stochastic optical
  reconstruction microscopy ({STORM})}.
\newblock \emph{\bibinfo{journal}{Nature Methods}}
  \textbf{\bibinfo{volume}{3}}, \bibinfo{pages}{793--796}
  (\bibinfo{year}{2006}).

\bibitem{betzig2006imaging}
\bibinfo{author}{Betzig, E.} \emph{et~al.}
\newblock \bibinfo{title}{Imaging intracellular fluorescent proteins at
  nanometer resolution}.
\newblock \emph{\bibinfo{journal}{Science}} \textbf{\bibinfo{volume}{313}},
  \bibinfo{pages}{1642--1645} (\bibinfo{year}{2006}).

\bibitem{dertinger2009fast}
\bibinfo{author}{Dertinger, T.}, \bibinfo{author}{Colyer, R.},
  \bibinfo{author}{Iyer, G.}, \bibinfo{author}{Weiss, S.} \&
  \bibinfo{author}{Enderlein, J.}
\newblock \bibinfo{title}{Fast, background-free, 3d super-resolution optical
  fluctuation imaging ({SOFI})}.
\newblock \emph{\bibinfo{journal}{Proceedings of the National Academy of
  Sciences}} \textbf{\bibinfo{volume}{106}}, \bibinfo{pages}{22287--22292}
  (\bibinfo{year}{2009}).

\bibitem{SandoghdarReview15}
\bibinfo{author}{Weisenburger, S.} \& \bibinfo{author}{Sandoghdar, V.}
\newblock \bibinfo{title}{Light microscopy: an ongoing contemporary
  revolution}.
\newblock \emph{\bibinfo{journal}{Contemporary Physics}}
  \textbf{\bibinfo{volume}{56}}, \bibinfo{pages}{123--143}
  (\bibinfo{year}{2015}).

\bibitem{localization_review_NMeth2014}
\bibinfo{author}{Deschout, H.} \emph{et~al.}
\newblock \bibinfo{title}{Precisely and accurately localizing single emitters
  in fluorescence microscopy}.
\newblock \emph{\bibinfo{journal}{Nature Methods}}
  \textbf{\bibinfo{volume}{11}}, \bibinfo{pages}{253--266}
  (\bibinfo{year}{2014}).

\bibitem{Small_biophys2008}
\bibinfo{author}{Small, A.~R.}
\newblock \bibinfo{title}{Theoretical limits on errors and acquisition rates in
  localizing switchable fluorophores}.
\newblock \emph{\bibinfo{journal}{Biophysical Journal}}
  \textbf{\bibinfo{volume}{96}}, \bibinfo{pages}{L16--L18}
  (\bibinfo{year}{2009}).

\bibitem{Huang2011_multi}
\bibinfo{author}{Huang, F.}, \bibinfo{author}{Schwartz, S.~L.},
  \bibinfo{author}{Byars, J.~M.} \& \bibinfo{author}{Lidke, K.~A.}
\newblock \bibinfo{title}{{Simultaneous multiple-emitter fitting for single
  molecule super-resolution imaging}}.
\newblock \emph{\bibinfo{journal}{Biomedical Optics Express}}
  \textbf{\bibinfo{volume}{2}}, \bibinfo{pages}{1377--1393}
  (\bibinfo{year}{2011}).

\bibitem{Zhu2012NatMehod}
\bibinfo{author}{Zhu, L.}, \bibinfo{author}{Zhang, W.},
  \bibinfo{author}{Elnatan, D.} \& \bibinfo{author}{Huang, B.}
\newblock \bibinfo{title}{{Faster STORM using compressed sensing}}.
\newblock \emph{\bibinfo{journal}{Nature Methods}}
  \textbf{\bibinfo{volume}{9}}, \bibinfo{pages}{721--723}
  (\bibinfo{year}{2012}).

\bibitem{sCMOS_Loc_2013}
\bibinfo{author}{Huang, F.} \emph{et~al.}
\newblock \bibinfo{title}{Video-rate nanoscopy using scmos camera-specific
  single-molecule localization algorithms}.
\newblock \emph{\bibinfo{journal}{Nature Methods}}
  \textbf{\bibinfo{volume}{10}}, \bibinfo{pages}{653--658}
  (\bibinfo{year}{2013}).

\bibitem{TonyICA_APL13}
\bibinfo{author}{Barsic, A.} \& \bibinfo{author}{Piestun, R.}
\newblock \bibinfo{title}{Super-resolution of dense nanoscale emitters beyond
  the diffraction limit using spatial and temporal information}.
\newblock \emph{\bibinfo{journal}{Applied Physics Letters}}
  \textbf{\bibinfo{volume}{102}}, \bibinfo{pages}{231103}
  (\bibinfo{year}{2013}).

\bibitem{wang2012palmer}
\bibinfo{author}{Wang, Y.}, \bibinfo{author}{Quan, T.}, \bibinfo{author}{Zeng,
  S.} \& \bibinfo{author}{Huang, Z.-L.}
\newblock \bibinfo{title}{{PALMER}: a method capable of parallel localization
  of multiple emitters for high-density localization microscopy}.
\newblock \emph{\bibinfo{journal}{Optics express}}
  \textbf{\bibinfo{volume}{20}}, \bibinfo{pages}{16039--16049}
  (\bibinfo{year}{2012}).

\bibitem{localization_algorithms_review_NMeth2014}
\bibinfo{author}{Small, A.} \& \bibinfo{author}{Stahlheber, S.}
\newblock \bibinfo{title}{Fluorophore localization algorithms for
  super-resolution microscopy}.
\newblock \emph{\bibinfo{journal}{Nature methods}}
  \textbf{\bibinfo{volume}{11}}, \bibinfo{pages}{267--279}
  (\bibinfo{year}{2014}).

\bibitem{OsipSR_PRA}
\bibinfo{author}{Schwartz, O.} \& \bibinfo{author}{Oron, D.}
\newblock \bibinfo{title}{Improved resolution in fluorescence microscopy using
  quantum correlations}.
\newblock \emph{\bibinfo{journal}{Phys. Rev. A}} \textbf{\bibinfo{volume}{85}},
  \bibinfo{pages}{033812} (\bibinfo{year}{2012}).

\bibitem{Tsang_PhysRevLett2016}
\bibinfo{author}{Nair, R.} \& \bibinfo{author}{Tsang, M.}
\newblock \bibinfo{title}{Far-field superresolution of thermal electromagnetic
  sources at the quantum limit}.
\newblock \emph{\bibinfo{journal}{Phys. Rev. Lett.}}
  \textbf{\bibinfo{volume}{117}}, \bibinfo{pages}{190801}
  (\bibinfo{year}{2016}).

\bibitem{CosmoPirandola_PRL2016}
\bibinfo{author}{Lupo, C.} \& \bibinfo{author}{Pirandola, S.}
\newblock \bibinfo{title}{Ultimate precision bound of quantum and subwavelength
  imaging}.
\newblock \emph{\bibinfo{journal}{Phys. Rev. Lett.}}
  \textbf{\bibinfo{volume}{117}}, \bibinfo{pages}{190802}
  (\bibinfo{year}{2016}).

\bibitem{OsipNanoletters2013}
\bibinfo{author}{Schwartz, O.} \emph{et~al.}
\newblock \bibinfo{title}{Superresolution microscopy with quantum emitters}.
\newblock \emph{\bibinfo{journal}{Nano Letters}} \textbf{\bibinfo{volume}{13}},
  \bibinfo{pages}{5832--5836} (\bibinfo{year}{2013}).

\bibitem{CuiPRL2013}
\bibinfo{author}{Cui, J.-M.}, \bibinfo{author}{Sun, F.-W.},
  \bibinfo{author}{Chen, X.-D.}, \bibinfo{author}{Gong, Z.-J.} \&
  \bibinfo{author}{Guo, G.-C.}
\newblock \bibinfo{title}{Quantum statistical imaging of particles without
  restriction of the diffraction limit}.
\newblock \emph{\bibinfo{journal}{Phys. Rev. Lett.}}
  \textbf{\bibinfo{volume}{110}}, \bibinfo{pages}{153901}
  (\bibinfo{year}{2013}).

\bibitem{COPS_CPC_2014}
\bibinfo{author}{Gru{\ss}mayer, K.~S.}, \bibinfo{author}{Kurz, A.} \&
  \bibinfo{author}{Herten, D.-P.}
\newblock \bibinfo{title}{Single-molecule studies on the label number
  distribution of fluorescent markers}.
\newblock \emph{\bibinfo{journal}{ChemPhysChem}} \textbf{\bibinfo{volume}{15}},
  \bibinfo{pages}{734--742} (\bibinfo{year}{2014}).

\bibitem{MappingSTED_NComm15}
\bibinfo{author}{Ta, H.} \emph{et~al.}
\newblock \bibinfo{title}{Mapping molecules in scanning far-field fluorescence
  nanoscopy}.
\newblock \emph{\bibinfo{journal}{Nature Communications}}
  \textbf{\bibinfo{volume}{6}}, \bibinfo{pages}{{7977}} (\bibinfo{year}{2015}).

\bibitem{AccuracityEMCCDsNMeth13}
\bibinfo{author}{Chao, J.}, \bibinfo{author}{Ram, S.}, \bibinfo{author}{Ward,
  E.~S.} \& \bibinfo{author}{Ober, R.~J.}
\newblock \bibinfo{title}{Ultrahigh accuracy imaging modality for
  super-localization microscopy}.
\newblock \emph{\bibinfo{journal}{Nature Methods}}
  \textbf{\bibinfo{volume}{10}}, \bibinfo{pages}{335--338}
  (\bibinfo{year}{2013}).

\bibitem{SPcams2014}
\bibinfo{author}{Krishnaswami, V.}, \bibinfo{author}{Van~Noorden, C.~J.},
  \bibinfo{author}{Manders, E.~M.} \& \bibinfo{author}{Hoebe, R.~A.}
\newblock \bibinfo{title}{Towards digital photon counting cameras for
  single-molecule optical nanoscopy}.
\newblock \emph{\bibinfo{journal}{Optical Nanoscopy}}
  \textbf{\bibinfo{volume}{3}}, \bibinfo{pages}{1} (\bibinfo{year}{2014}).

\bibitem{microlens2014}
\bibinfo{author}{Pavia, J.~M.}, \bibinfo{author}{Wolf, M.} \&
  \bibinfo{author}{Charbon, E.}
\newblock \bibinfo{title}{Measurement and modeling of microlenses fabricated on
  single-photon avalanche diode arrays for fill factor recovery}.
\newblock \emph{\bibinfo{journal}{Optics express}}
  \textbf{\bibinfo{volume}{22}}, \bibinfo{pages}{4202--4213}
  (\bibinfo{year}{2014}).

\bibitem{OrritPhotonSource2005}
\bibinfo{author}{Lounis, B.} \& \bibinfo{author}{Orrit, M.}
\newblock \bibinfo{title}{Single-photon sources}.
\newblock \emph{\bibinfo{journal}{Reports on Progress in Physics}}
  \textbf{\bibinfo{volume}{68}}, \bibinfo{pages}{{1129--1179}}
  (\bibinfo{year}{2005}).

\bibitem{IntroQuantOPt}
\bibinfo{author}{Gerry, C.~C.} \& \bibinfo{author}{Knight, P.~L.}
\newblock \emph{\bibinfo{title}{Introductory Quantum Optics}}
  (\bibinfo{publisher}{Cambridge University Press}, \bibinfo{year}{2005}).

\bibitem{mortensenNMeth2010}
\bibinfo{author}{Mortensen, K.~I.}, \bibinfo{author}{Churchman, L.~S.},
  \bibinfo{author}{Spudich, J.~A.} \& \bibinfo{author}{Flyvbjerg, H.}
\newblock \bibinfo{title}{Optimized localization analysis for single-molecule
  tracking and super-resolution microscopy}.
\newblock \emph{\bibinfo{journal}{Nature Methods}}
  \textbf{\bibinfo{volume}{7}}, \bibinfo{pages}{377--381}
  (\bibinfo{year}{2010}).

\bibitem{CovaGhioni2011}
\bibinfo{author}{Cova, S.~D.} \& \bibinfo{author}{Ghioni, M.}
\newblock \bibinfo{title}{Single-photon counting detectors}.
\newblock \emph{\bibinfo{journal}{Photonics Journal, IEEE}}
  \textbf{\bibinfo{volume}{3}}, \bibinfo{pages}{274--277}
  (\bibinfo{year}{2011}).

\bibitem{CharbonFishburn2011}
\bibinfo{author}{Charbon, E.} \& \bibinfo{author}{Fishburn, M.~W.}
\newblock \bibinfo{title}{Monolithic single-photon avalanche diodes: Spads}.
\newblock In \emph{\bibinfo{booktitle}{Single-Photon Imaging}},
  \bibinfo{pages}{123--157} (\bibinfo{publisher}{Springer},
  \bibinfo{year}{2011}).

\bibitem{ISM_PRL10}
\bibinfo{author}{M\"uller, C.~B.} \& \bibinfo{author}{Enderlein, J.}
\newblock \bibinfo{title}{Image scanning microscopy}.
\newblock \emph{\bibinfo{journal}{Phys. Rev. Lett.}}
  \textbf{\bibinfo{volume}{104}}, \bibinfo{pages}{198101}
  (\bibinfo{year}{2010}).

\bibitem{ConicalSuperRes13}
\bibinfo{author}{Rosen, S.}, \bibinfo{author}{Sirat, G.~Y.},
  \bibinfo{author}{Ilan, H.} \& \bibinfo{author}{Agranat, A.~J.}
\newblock \bibinfo{title}{A sub wavelength localization scheme in optical
  imaging using conical diffraction}.
\newblock \emph{\bibinfo{journal}{Optics Express}}
  \textbf{\bibinfo{volume}{21}}, \bibinfo{pages}{10133--10138}
  (\bibinfo{year}{2013}).

\bibitem{Airyscan2015}
\bibinfo{author}{Huff, J.}
\newblock \bibinfo{title}{The {Airyscan} detector from {ZEISS}: confocal
  imaging with improved signal-to-noise ratio and super-resolution}.
\newblock \emph{\bibinfo{journal}{Nature Methods}}
  \textbf{\bibinfo{volume}{12}}, \bibinfo{pages}{12} (\bibinfo{year}{2015}).

\bibitem{DyeAntibunchingNature2000}
\bibinfo{author}{Lounis, B.} \& \bibinfo{author}{Moerner, W.~E.}
\newblock \bibinfo{title}{Single photons on demand from a single molecule at
  room temperature}.
\newblock \emph{\bibinfo{journal}{Nature}} \textbf{\bibinfo{volume}{407}},
  \bibinfo{pages}{491--493} (\bibinfo{year}{2000}).

\bibitem{Review-SRprobes08}
\bibinfo{author}{Fern{\'a}ndez-Su{\'a}rez, M.} \& \bibinfo{author}{Ting, A.~Y.}
\newblock \bibinfo{title}{Fluorescent probes for super-resolution imaging in
  living cells}.
\newblock \emph{\bibinfo{journal}{Nature Reviews Molecular Cell Biology}}
  \textbf{\bibinfo{volume}{9}}, \bibinfo{pages}{929--943}
  (\bibinfo{year}{2008}).

\bibitem{singleDyesAndProteinsLocACSNano12}
\bibinfo{author}{Han, J.~J.}, \bibinfo{author}{Kiss, C.},
  \bibinfo{author}{Bradbury, A.~R.} \& \bibinfo{author}{Werner, J.~H.}
\newblock \bibinfo{title}{Time-resolved, confocal single-molecule tracking of
  individual organic dyes and fluorescent proteins in three dimensions}.
\newblock \emph{\bibinfo{journal}{ACS Nano}} \textbf{\bibinfo{volume}{6}},
  \bibinfo{pages}{8922--8932} (\bibinfo{year}{2012}).

\bibitem{donnert2007major}
\bibinfo{author}{Donnert, G.}, \bibinfo{author}{Eggeling, C.} \&
  \bibinfo{author}{Hell, S.~W.}
\newblock \bibinfo{title}{Major signal increase in fluorescence microscopy
  through dark-state relaxation}.
\newblock \emph{\bibinfo{journal}{Nature Methods}}
  \textbf{\bibinfo{volume}{4}}, \bibinfo{pages}{81--86} (\bibinfo{year}{2007}).

\bibitem{jacques2008enhancing}
\bibinfo{author}{Jacques, V.} \emph{et~al.}
\newblock \bibinfo{title}{Enhancing single-molecule photostability by optical
  feedback from quantum jump detection}.
\newblock \emph{\bibinfo{journal}{Applied Physics Letters}}
  \textbf{\bibinfo{volume}{93}}, \bibinfo{pages}{203307}
  (\bibinfo{year}{2008}).

\bibitem{FALCON2014}
\bibinfo{author}{Min, J.} \emph{et~al.}
\newblock \bibinfo{title}{{FALCON}: fast and unbiased reconstruction of
  high-density super-resolution microscopy data}.
\newblock \emph{\bibinfo{journal}{Scientific Reports}}
  \textbf{\bibinfo{volume}{4}}, \bibinfo{pages}{4577} (\bibinfo{year}{2014}).

\bibitem{SparseLoc2016}
\bibinfo{author}{Siewert, S.} \emph{et~al.}
\newblock \bibinfo{title}{Sparse deconvolution of high-density super-resolution
  images}.
\newblock \emph{\bibinfo{journal}{Scientific Reports}}
  \textbf{\bibinfo{volume}{6}}, \bibinfo{pages}{21413} (\bibinfo{year}{2016}).

\bibitem{QC_SPADS_PRL2014}
\bibinfo{author}{Gatto~Monticone, D.} \emph{et~al.}
\newblock \bibinfo{title}{Beating the {Abbe} diffraction limit in confocal
  microscopy via nonclassical photon statistics}.
\newblock \emph{\bibinfo{journal}{Phys. Rev. Lett.}}
  \textbf{\bibinfo{volume}{113}}, \bibinfo{pages}{143602}
  (\bibinfo{year}{2014}).

\bibitem{rawdata}
\bibinfo{author}{Tenne, R.}, \bibinfo{author}{Israel, Y.},
  \bibinfo{author}{Oron, D.} \& \bibinfo{author}{Silberberg, Y.}
\newblock \bibinfo{title}{{ExampleData2.zip}}  (\bibinfo{year}{2017}).
\newblock
  \urlprefix\url{https://figshare.com/articles/ExampleData2_zip/4588723}.

\end{thebibliography}

\end{document}